\documentclass[aps,prc,10pt,showpacs,showkeys,onecolumn,superscriptaddress,groupedaddress]{revtex4-1}
\usepackage{amssymb}
\usepackage[english]{babel}
\usepackage{indentfirst}
\usepackage{amsxtra}
\usepackage{amsmath}
\usepackage{supertabular}
\usepackage{multirow}
\usepackage [mathcal]{eucal}
\usepackage{graphicx}
\usepackage{graphics}

\def\beq{\begin{equation}} 
\def\eeq{\end{equation}}
\begin{document}
\noindent
\title{Neutron density distribution and neutron skin thickness of $^{208}$Pb }
\author{Andrea Meucci}
\author{Matteo Vorabbi}
\author{Carlotta Giusti}
\affiliation{Dipartimento di Fisica,  
Universit\`a degli Studi di Pavia and \\
INFN, Sezione di Pavia,  Via A. Bassi 6, I-27100 Pavia, Italy}
\author{Paolo Finelli}
\affiliation{Dipartimento di Fisica e Astronomia, 
Universit\`{a} degli Studi di Bologna and \\
INFN,
Sezione di Bologna, Via Irnerio 46, I-40126 Bologna, Italy}
\date{\today}

\begin{abstract} 
We present and discuss numerical predictions for the neutron 
density distribution of  $^{208}$Pb using
 various non-relativistic and relativistic mean-field  models for the nuclear structure.
Our results are compared with the very recent  pion photoproduction data from Mainz. 
The parity-violating asymmetry parameter  for elastic electron scattering 
at the kinematics of the  PREX experiment at JLab
and the neutron skin thickness are compared with the available data. 
We consider also the  dependence between the neutron skin and the 
parameters of the expansion of the symmetry energy.
\end{abstract}

\pacs{21.60.-n; 21.60.Jz; 25.30.Bf; 21.65.Ef}

\maketitle

\section{Introduction}

An accurate description of matter distribution in nuclei is a longstanding
problem in modern nuclear physics that has a wide impact on our
understanding of nuclear structure. Whereas the charge distribution
has been measured  with high accuracy using electron-nucleus elastic scattering, 
so that the charge radii are usually known with uncertainties lower 
than $1\%$ \cite{Angeli2004185,Angeli201369}, our knowledge of
neutron distribution is considerably less precise.
Several experiments of neutron radius have been carried out 
in recent years \cite{PhysRevC.76.014311,PhysRevC.76.034305,PhysRevC.82.044611},
but the use of hadronic probes  
produces uncertainties in the experimental results due to the assumptions of the
models required to deal with the complexity of the strong interaction.
An accurate and model independent probe of neutron distributions is provided by 
parity-violating electron scattering (PVES):  the parity-violating 
asymmetry $A_{pv}$,  i.e.,  the difference between the cross sections 
for the scattering of electrons longitudinally polarized parallel and 
antiparallel  to their momentum, represents an almost direct measurement of the 
Fourier transform of the neutron density 
\cite{Donnelly1989589,PhysRevC.61.064307}
that is free from most strong interaction uncertainties.

The PREX Collaboration \cite{Abrahamyan:2012gp} at JLab  
used parity-violating electron scattering (PVES) to study
the neutron distribution of $^{208}$Pb and provided us with the 
first determination of the neutron radius through an electroweak probe that gives
$R_{skin} = 0.33 ^{+0.16}_{-0.18}$ fm for the neutron skin thickness. Although the total error
is large, the PREX method is very interesting and future higher statistics 
data are expected to reduce the uncertainty \cite{prex2}.
The CREX Collaboration at JLab has made a successful 
proposal to measure the neutron radius of $^{48}$Ca using  
PVES with a goal of $\pm 0.02$ fm in accuracy \cite{crex}.
Recently, in a measurement of the coherent $\pi^0$ photoproduction from 
$^{208}$Pb at Mainz \cite{mainz-neutronskin},
the shape of the neutron distribution has been found to be $20\%$ more diffuse
than the  charge distribution and the neutron skin thickness is
$R_{skin} = 0.15\pm 0.03$ (stat) $^{+0.01}_{-0.03}$ (syst) fm. This value is compatible with
previous independent measurements, i.e., proton elastic 
scattering \cite{PhysRevC.82.044611,PhysRevC.49.2118},
x-ray cascade of antiprotonic atoms \cite{PhysRevC.76.014311,PhysRevC.76.034305}, 
anti-analog giant dipole resonances \cite{Krasznahorkay:2013lga,
Krasznahorkay2013428,Krasznahorkay:2013ozl},  giant quadrupole 
resonances \cite{PhysRevC.87.034301}, pigmy dipole resonance \cite{PhysRevC.76.051603,
PhysRevC.81.041301,PhysRevC.84.021302,PhysRevC.85.041304}, 
electric dipole polarizability \cite{PhysRevLett.107.062502,PhysRevLett.107.062502}
or pionic probes \cite{Friedman201246}.

The neutron skin of $^{208}$Pb has important implications for 
astrophysics \cite{brown,PhysRevLett.86.5647,PhysRevC.64.062802}, owing to its
strong correlation with the pressure of neutron matter at densities near 
$0.1$ fm$^{-3}$.
The larger the pressure of neutron matter, the thicker is the skin as neutrons
are pushed out against surface tension. 
The same pressure supports neutron stars against gravity, therefore 
correlations between neutron skins of neutron-rich nuclei and 
various neutron star properties are naturally expected \cite{PhysRevC.86.015802,lattimer14}.
In addition, the magnitude of $R_{skin}$ in heavy nuclei
provides very interesting information  on the nature of 3-body forces in nuclei,
nuclear drip lines and collective nuclear excitations, as well as
heavy-ion collisions. A recent review of experimental measurements of $R_{skin}$ and their
theoretical implications can be found in \cite{PhysRevC.86.015803}.

In this work we present and discuss numerical predictions for the neutron 
density distribution of  $^{208}$Pb.
In \cite{esotici1,esotici2} we have already considered the evolution
of the charge density
distribution and of the proton wave functions along different isotopic chains.
In \cite{isotoni} we have extended our study to isotonic chains.
In \cite{esotici2} we have already compared our calculations 
for the asymmetry parameter $A_{pv}$ using the relativistic DDME2 interaction
with the results of the first run of PREX 
on $^{208}$Pb and we have provided numerical predictions for the 
future experiment CREX on $^{48}$Ca. In addition, we have studied the behavior 
of $A_{pv}$ along oxygen and calcium isotopic chains \cite{esotici2}. 
In this paper we extend the work undertaken in \cite{esotici2} comparing 
results obtained with  different non-relativistic and relativistic
model interactions. Our results are compared with the recent $(\gamma , \pi^0)$
data from Mainz and with the data of the  PREX experiment. 
In addition, we consider also the correlations 
between the neutron skin and the slope and curvature coefficients of the 
nuclear symmetry energy.

\section{Neutron distribution of $^{208}$Pb } \label{sec2}
\begin{table}[t]
\begin{center}
\begin{ruledtabular}
\begin{tabular}{ccc}
\hline
  Interaction  &    $R$ & $a$  \\
 &   [fm] & [fm] \\
\hline
          L2 \cite{cnp} &    6.832 (9)  &  0.522 (8)\\
          NL3 \cite{Lalazissis:1996rd}  &    6.902 (7)  &  0.556 (6)\\
          NL3-II \cite{Lalazissis:1996rd}  &    6.888 (7)  &  0.557 (6)\\
          NL-SH \cite{Sharma1993377}  &    6.895 (9)  &  0.527 (8)\\
          DDME1 \cite{PhysRevC.66.024306} &    6.770 (9)  &  0.574 (7)\\
          DDME2 \cite{PhysRevC.71.024312} &    6.758 (9)  &  0.570 (8)\\
          PKDD \cite{PhysRevC.69.034319} &  6.832 (9) & 0.562 (7) \\
          DDPC1 \cite{PhysRevC.78.034318} & 6.783 (7) & 0.573 (6) \\ 
          PC-F1 \cite{pointc} &    6.903 (7)  &  0.566 (6)\\
          PC-F2 \cite{pointc} &    6.900 (7)  &  0.566 (6)\\
          PC-F4 \cite{pointc} &    6.899 (6)  &  0.567 (5)\\
          D1S \cite{PhysRevC.21.1568} &   6.697 (21)   & 0.575 (18) \\
          SIII \cite{Beiner197529} &   6.854 (4)   & 0.528 (3) \\
           SKM* \cite{Bartel198279} &   6.746 (4)   & 0.583 (3) \\
            SLY4 \cite{Chabanat1998231} &   6.752 (4)   & 0.582 (6) \\
           SLY5 \cite{Chabanat1998231} &   6.744 (7)   & 0.582 (6) \\
           SIII (mod) \cite{PhysRevC.88.054328} &   6.860 (4)   & 0.537 (3) \\
           SLY5 (mod) \cite{PhysRevC.88.054328} &   6.754 (7)   & 0.595 (6) \\
\hline
\end{tabular}
\end{ruledtabular}
\caption{\small Predictions for the half-height radius $R$ and diffuseness $a$ of $^{208}$Pb
from  various nuclear structure calculations. 
In parentheses the error on the last significant digit. 
The experimental data of \cite{mainz-neutronskin} are 
$R = 6.70 \pm 0.03$ fm and $a  = 0.55\pm  0.01 $ (stat) 
$^{+0.02}_{-0.03}$ (syst) fm.
}  
\label{tab:fiteng}
\end{center} 
\end{table}
\begin{figure}
	\centering \vskip 1mm
		\includegraphics[scale=0.34, bb= -12 25 720 527, clip]{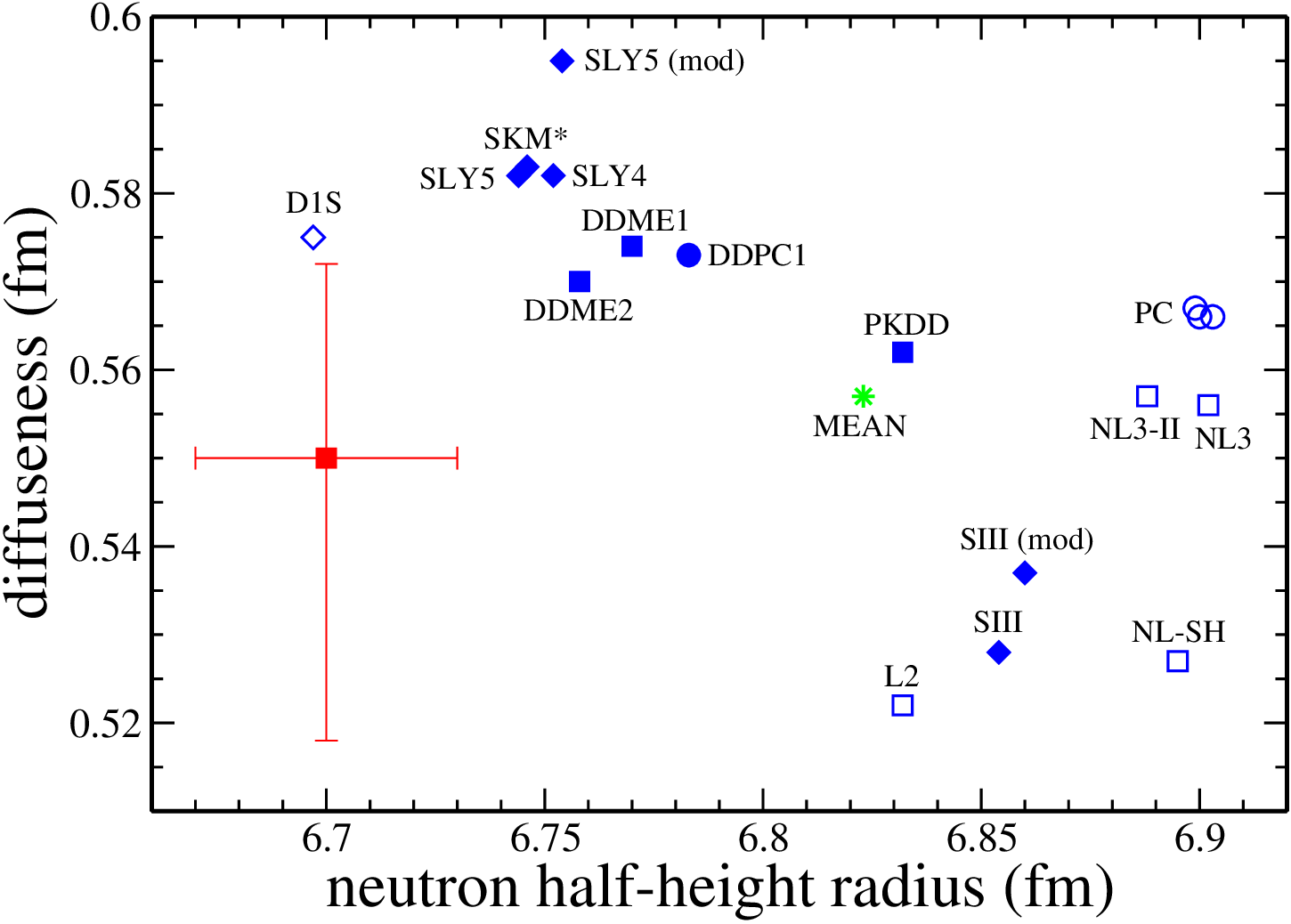}
\caption{ (Color online) The half-height radius plotted versus the diffuseness
for $^{208}$Pb.
The red square shows the experimental data of  \cite{mainz-neutronskin} 
 with statistical and systematic errors. 
}
\label{fig:r_a}
\end{figure}

The best description of heavy nuclei, at the moment, relies on energy density functionals 
in terms of 
effective interactions calibrated on the bulk properties of a limited set of nuclei. 
The isoscalar part of the interaction is usually constrained by reproducing 
binding energies and charge radii ($^{208}$Pb is usually included in fit protocol) where 
the isospin-dependent part of the interaction is mainly constrained reproducing some 
{\it ab-initio} equation of state (EOS) for neutron matter, like the 
Akmal-Friedmann-Pandharipande EOS \cite{PhysRevC.58.1804}, or the empirical value of 
the asymmetry energy at the saturation point. 
So far, theoretical calculations based on realistic potentials are limited to
medium-light nuclei, even if new approaches based on 
renormalization group potentials look very promising \cite{Binder:2013xaa}.
In this work we consider different non-relativistic  and relativistic mean-field (RMF) 
models and compare their predictions for the neutron distribution of $^{208}$Pb. 
The details 
of the mean-field approaches we have adopted in our investigation are presented in 
various publications, for instance in \cite{brink,Decharge:1979fa,
PhysRevC.21.1568,Sharma1993377,Ring:1996,
Serot:1997,Lalazissis:1996rd,PhysRevC.66.024306,PhysRevC.69.034319,
PhysRevC.71.024312,PhysRevC.78.034318,pointc,Beiner197529,Bartel198279,
Chabanat1998231,Chabanat1998231,PhysRevC.88.054328}. We do not repeat
here the derivation of the various expressions used in our
calculations but we refer the readers to the original papers. Our strategy is to explore all
variants of density functional approaches in terms of covariant 
(Walecka type) {\it vs.} non-covariant (Skyrme and Gogny) descriptions, finite 
range {\it vs.} contact interactions and non-linear {\it vs.} density dependent couplings.

We have checked that the different forces adopted for our calculations give 
some differences in the neutron single-particle levels  
around the Fermi surface in $^{208}$Pb, but do not produce significant
inversions in the energy levels. 
The levels above the $N = 126$ shell closure are unoccupied. 

Generally, the nucleon  distributions are parameterized as a single 
 symmetrised two-parameter Fermi distribution  (2pF)  \cite{PhysRevC.36.1105} 
 with half-height radius $R$ and
diffuseness $a$. The analysis of the $(\gamma, \pi^0)$ cross sections data from Mainz gives
$R = 6.70 \pm 0.03$ fm and $a  = 0.55\pm  0.01 $ (stat) 
$^{+0.02}_{-0.03}$ (syst) fm \cite{mainz-neutronskin} and suggests that the
neutron distribution of $^{208}$Pb is  $\approx 20\%$ more
diffuse than the charge distribution and that the neutron skin of lead is of partial 
halo type.

In Table \ref{tab:fiteng} we report our results for 
the half-heigth radius and diffuseness parameter of the 2pF neutron density 
distributions extracted from the different models. In Fig. \ref{fig:r_a} these results are directly compared
with the experimental data for $R$ and $a$. Neither the nonrelativistic Gogny
and Skyrme  nor the RMF
models are able to simultaneously reproduce the experimental data for both $R$
and $a$. The finite-range Gogny D1S interaction reproduces $R$ but slightly 
overestimates $a$. 
The Skyrme interaction parametrizations (those starting with S) generally 
give similar results for $R$ and $a$ that reproduce the experimental value of 
$R$ within two standard deviations and overestimate $a$, but the SIII interactions 
that reproduce the
experimental value of $a$ and overestimate $R$. 
The RMF models that include nonlinear self-interaction meson couplings (those
starting with NL)  reproduce the diffuseness but overestimate the radius
over three standard deviations. 
These results are consistent with the observation that the 
mixed isoscalar-isovector coupling
terms  in the Lagrangian densities should be taken into account to significantly
change the neutron radii \cite{PhysRevC.64.062802,fsugold}.
The RMF models with point-coupling interaction (those with PC), i.e., where
the zero-range point-coupling interaction is used instead of
the meson exchange, give almost coincident results that overestimate  $R$ and reproduce $a$.
The  relativistic functionals with density-dependent vertex functions (those
starting with DD) agree with the experimental diffuseness and reproduce the radius within two 
standard deviations. The density-dependent 
PKDD model overestimates $R$ and reproduce $a$.

To obtain a simple model of the neutron density distributions, we have evaluated 
the weighted average  parameters of the 2pF profiles extracted from the results 
of the different models and have obtained $R_{mean} = 6.822 \pm 0.001$ fm and $a_{mean} = 0.558 \pm 0.001$ fm: the 
surface diffuseness is in fair agreement with the Mainz data but the radius is 
a bit larger. 
This \lq\lq weighted\rq\rq\ result is plotted with the green square (MEAN)
in  Fig. \ref{fig:r_a}. To be more confident, we have checked that the
2pF profile of the charge distribution obtained with this weighted average  
procedure is able to satisfactorily reproduce the experimental data of elastic 
electron scattering cross sections off $^{208}$Pb.

\subsection{Comparison with PREX} \label{prex}

\begin{figure}
	\centering \vskip 1mm
		\includegraphics[scale=0.34, bb= -12 25 720 527, clip]{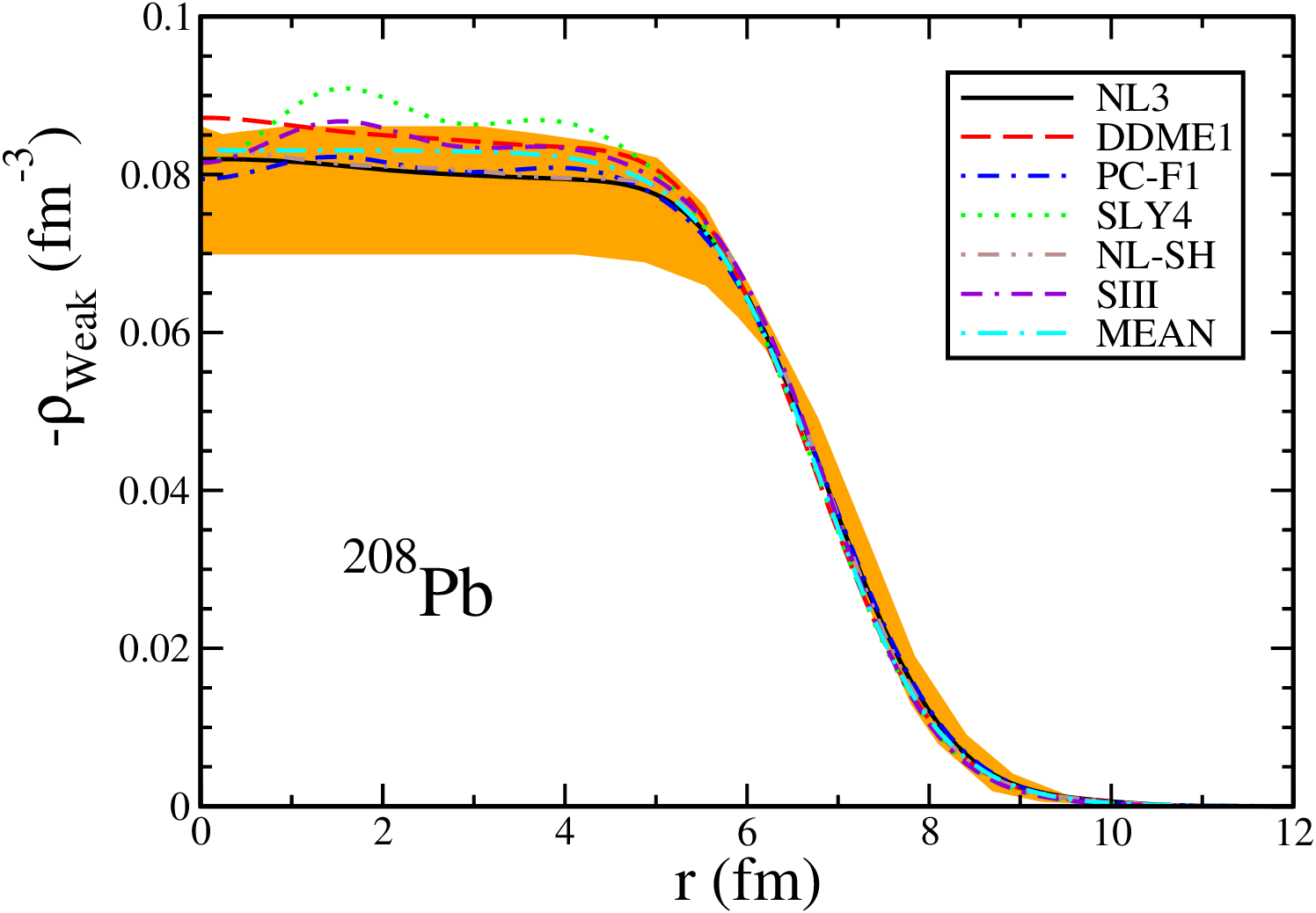}
\vskip 1mm
\caption{ (Color online) 
Theoretical weak charge density  in 
comparison with the experimental error band as determined in 
Ref. \cite{PhysRevC.85.032501} for $^{208}$Pb with the kinematics
of the PREX experiment.
}	
\label{fig:rhoweak}
\end{figure}
\begin{figure}
	\centering \vskip 6mm
		\includegraphics[scale=0.34, bb= -12 22 720 527, clip]{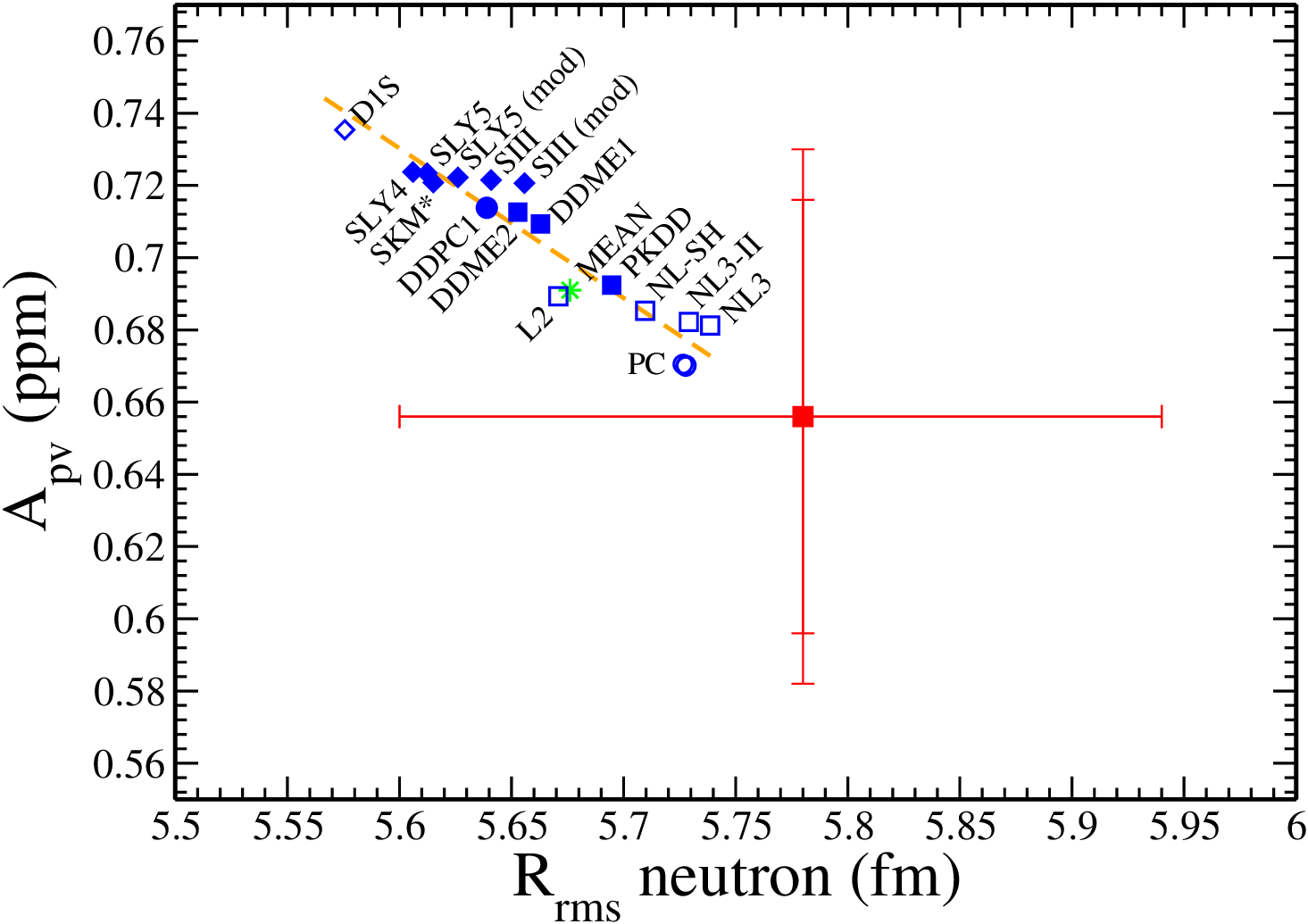}
\vskip 1mm
\caption{ (Color online) 
Parity-violating asymmetry at the kinematics of the PREX experiments versus
 the neutron rms radius for $^{208}$Pb. The dashed orange line is a linear fit
 of the correlation between the neutron rms radius and $A_{pv}$.
 The red square shows the experimental data from PREX \cite{Abrahamyan:2012gp}
 with statistical and systematic errors. 
}	
\label{fig:apv_rneu}
\end{figure}
\begin{figure}
	\centering \vskip 1mm
		\includegraphics[scale=0.34, bb= -12 25 720 527, clip]{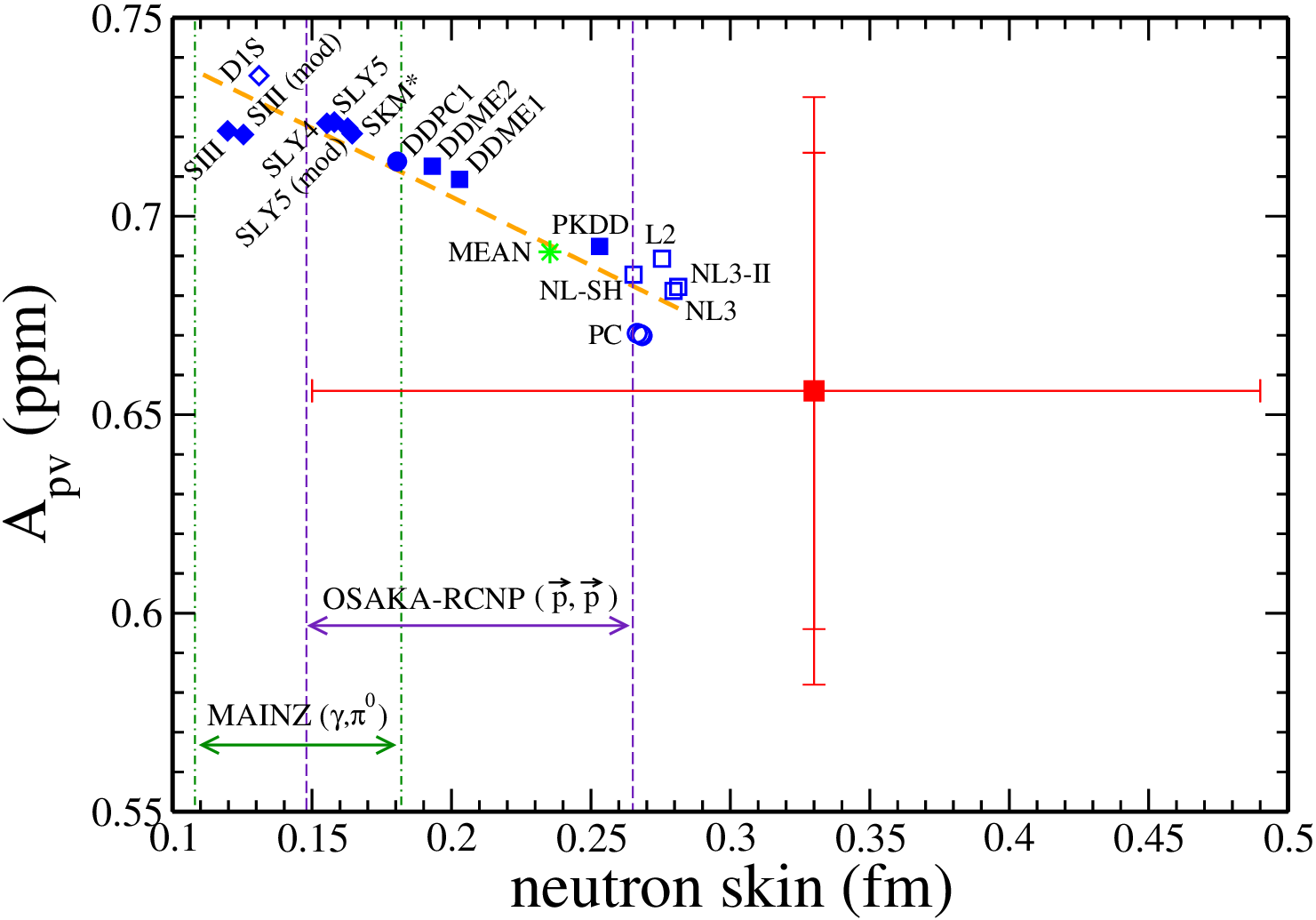}
\vskip 1mm
\caption{ (Color online) 
Parity-violating asymmetry at the kinematics of the PREX experiments versus
 the neutron skin for $^{208}$Pb.  The red square shows the experimental 
 data from PREX \cite{Abrahamyan:2012gp} with statistical and systematic errors. 
 The dashed orange line is a linear fit
 of the correlation between the neutron skin and $A_{pv}$.
 The vertical solid green lines show the constraints
 on $R_{skin}$
 from Mainz $(\gamma , \pi^{0})$ measurements \cite{mainz-neutronskin}.
 The vertical purple dashed lines  show the constraints on $R_{skin}$
 from Osaka polarized proton elastic scattering measurements \cite{PhysRevC.82.044611}.
}
\label{fig:apv_skin}
\end{figure}

The  parity-violating asymmetry parameter $A_{pv}$ is defined as the difference between 
the cross sections for the elastic scattering of 
electrons longitudinally polarized parallel and antiparallel to their momentum.  
$A_{pv}$ is proportional to the weak form
factor and, in Born approximation, it is very close to the
Fourier transform of the neutron density.

In Fig. \ref{fig:rhoweak} we show our theoretical predictions for the weak charge 
density $(-\rho_W)$  that has been deduced from the weak charge form 
factor \cite{PhysRevC.85.032501,Reinhard:2013fpa}. The error band (shaded area) 
represents the incoherent 
sum of experimental and model errors. Owing to the fact that the $Z^0$ boson couples mainly
with the neutron, $\rho_W$ depends essentially on the neutron distribution. 
Our predictions for different interactions are  in rather good agreement with the
empirical data. In addition, the weak distribution evaluated using the 2pF functions
for the proton
and neutron density distributions with weighted average parameters 
is also in good agreement with the data.

The PREX Collaboration measured the parity-violating asymmetry 
parameter $A_{pv}$  averaged over the experimental acceptance 
function $\epsilon(\theta)$ \cite{acceptance}
\begin{equation}
\langle A_{pv} \rangle = \dfrac{\int {\rm d}\theta~ \sin \theta A_{pv}(\theta) 
\dfrac{d\sigma}{d\Omega} \  \epsilon(\theta)}
{\int {\rm d}\theta~ \sin \theta \ \dfrac{d\sigma}{d\Omega} \ \epsilon(\theta)} \ ,
\end{equation}
where $A_{pv}(\theta)$ and $d\sigma / d\Omega$ are the 
asymmetry and the differential cross section at the scattering angle $\theta$.
 The charge radius of $^{208}$Pb
is very well known \cite{Angeli2004185,Angeli201369}; therefore, the empirical estimate 
$A_{pv} = 0.656 \pm 0.060$(stat) $\pm 0.014$(syst) ppm can be 
related to the neutron radius and the neutron rms radius 
results  $R_{n} = 5.78 ^{+0.16}_{-0.18}$ fm that
implies that the neutron skin thickness is $R_{skin} = 0.33 ^{+0.16}_{-0.18}$ fm. 

The results for the parity-violating asymmetry 
$A_{pv}$ versus the neutron rms radius for different models are 
displayed in Fig. \ref{fig:apv_rneu}. 
The result with the 2pF functions
for the  density distributions with averaged parameters is also in good 
agreement with the data. It is interesting to observe that there is a
linear correlation between $A_{pv}$ and the neutron radius as well as the 
neutron skin \cite{PhysRevC.81.051303}. Our results in  
Fig. \ref{fig:apv_rneu} are in accordance with this observation. Owing to the large
experimental uncertainties, our theoretical predictions are in agreement
with the data but they all predict a smaller radius than the central value of $5.78$ fm.
To obtain a significantly larger $R_{n}$ and a smaller $A_{pv}$ 
the Lagrangian density should contain also the 
mixed isoscalar-isovector coupling
term as described in \cite{Ban:2010wx}. 
We observe, however, that a large neutron radius 
is not in agreement with other experimental measurements \cite{PhysRevC.86.015803}.

In Fig. \ref{fig:apv_skin} we present the results for the parity-violating
asymmetry versus the neutron skin predicted by the different models. 
Owing to the fact that the neutron skin is highly correlated 
with the neutron radius, these results are similar to those in  Fig. \ref{fig:apv_rneu}.
 The constraints on the neutron skin
 from Mainz $(\gamma , \pi^{0})$ \cite{mainz-neutronskin}, 
 as well as those from Osaka polarized proton elastic scattering 
 measurements at proton energy 
$\varepsilon = 295$ MeV \cite{PhysRevC.82.044611}, are displayed
 for a comparison. 
 Although all the predictions  in  Fig. \ref{fig:apv_rneu} and 
 \ref{fig:apv_skin} are compatible with the PREX results, the large error bars
 prevent us from discriminating among some of them. Other $R_{skin}$ measurements
 are more precise and seem to rule out models with either very small  
 or very large  neutron skins. However, a careful analysis of all available data 
 in \cite{PhysRevLett.111.162501} demonstrates that it is still premature
 to rule out  the existence of a thick neutron skin in $^{208}$Pb.


\subsection{Neutron skin and the symmetry energy at saturation density}
\label{symm}

Around the nuclear matter saturation density  $\rho_0$ the nuclear
symmetry energy can be expanded to second order in density as
\beq
e_{sym}(\rho) \simeq e_{sym}(\rho_0) + \frac{L}{3} \left( \frac{\rho-\rho_0}{\rho} \right) +
 \frac{K_{sym}}{18} \left( \frac{\rho-\rho_0}{\rho} \right) ^2 \ . \label{eq:esym}
\eeq
The coefficient of the linear term of the expansion
is directly related to the energy of pure neutron matter at $\rho_0$ and it is defined as 
\beq
L =  3 \rho_0 \
\left. \frac{\partial e_{sym}(\rho)} {\partial \rho} \right|_{\rho=\rho_0}  \ ,
\label{eq:L}
\eeq
and the curvature parameter is
\beq
K_{sym} = 9 {\rho_0}^2 \
\left. \frac{\partial ^2 e_{sym}(\rho)} {\partial \rho^2}  \right|_{\rho=\rho_0}  \ .
\label{eq:K}
\eeq

Owing to the fact that the thickness of the neutron skin results from an interplay between
the surface tension and the  gradient of the 
symmetry energy between the surface and the center of the nucleus, 
there is a well-established linear dependence between the neutron skin and $L$ that is
usually adopted to constrain the density derivative of the symmetry energy 
\cite{PhysRevC.68.064307,PhysRevC.72.064309,PhysRevC.79.057301,
PhysRevLett.102.122502,PhysRevC.80.024316,PhysRevC.82.024321,
PhysRevLett.106.252501,PhysRevLett.109.262501,Zhang2013234}.

\begin{figure}
	\centering \vskip 1mm
		\includegraphics[scale=0.34, bb= -12 32 720 527, clip]{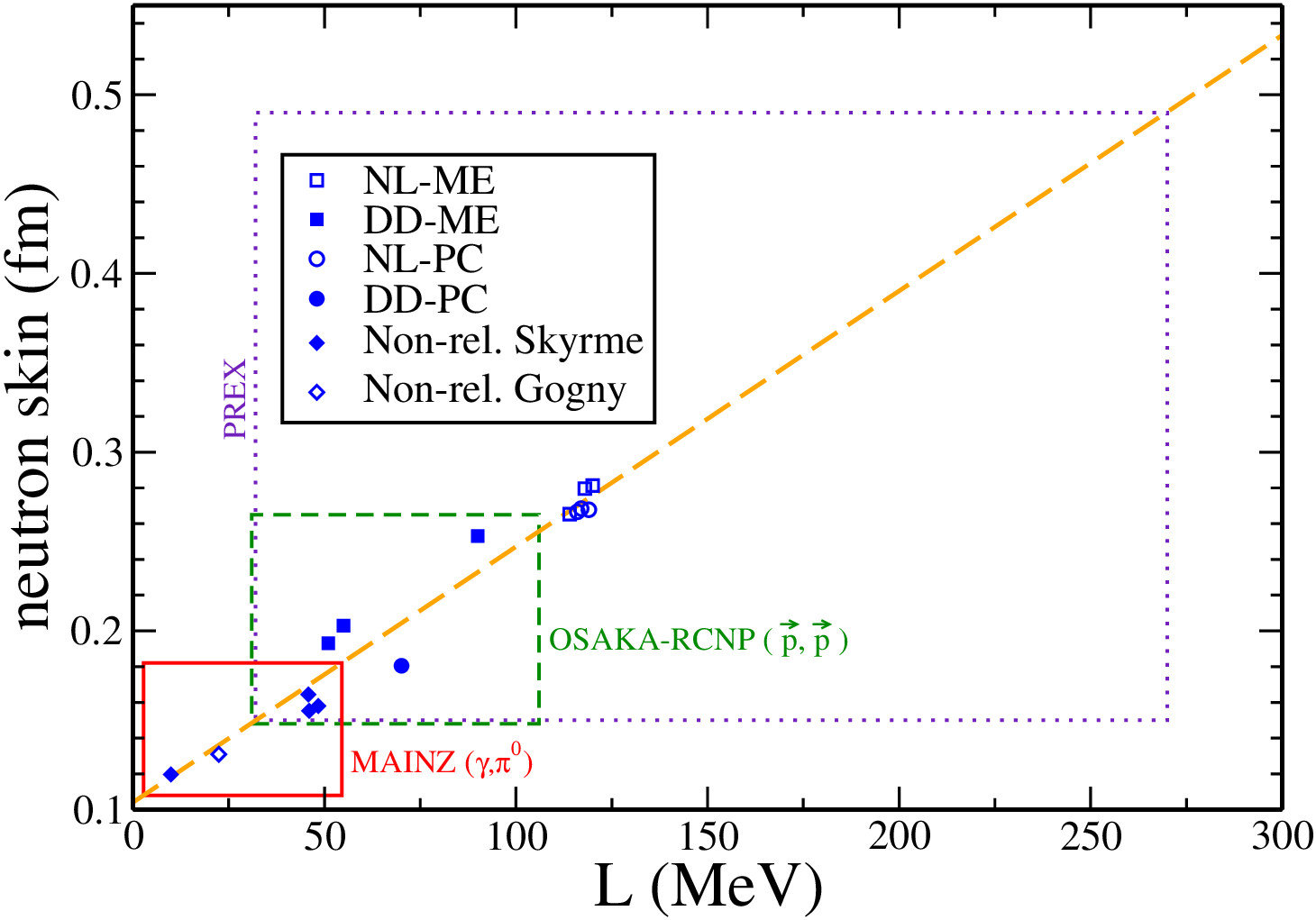}
\caption{ (Color online) 
Neutron skin for $^{208}$Pb versus
 the slope of the symmetry energy $L$ (the orange dashed line represents the best fit correlation).  The purple dotted lines are the 
 $1 \sigma$ contours extacted from PREX 
 \cite{Abrahamyan:2012gp}. The red solid lines show the constraints
 from Mainz $(\gamma , \pi^{0})$ measurements \cite{mainz-neutronskin} and
 the green dotted lines  show the constraints on $R_{skin}$
 from Osaka polarized proton elastic scattering measurements \cite{PhysRevC.82.044611}.
}	
\label{fig:l_skin}
\end{figure}

In Fig. \ref{fig:l_skin} we present the correlation of the neutron skin thickness of
 $^{208}$Pb versus $L$.
 The nonrelativistic Gogny and 
Skyrme interactions have soft symmetry energies $(L \leq 50$ MeV), while most relativistic
nuclear interactions lead to stiff symmetry energies ($L \geq 100$ MeV). The
inclusion of the  mixed isoscalar-isovector coupling
terms  in the Lagrangian densities produces a softer symmetry energy 
\cite{PhysRevC.64.062802,fsugold}. 
The  point-coupling interactions give  stiff  symmetry 
energy ($L \approx 120$ MeV) but the density-dependent interaction 
DDPC1 produces $L \approx 70$ MeV. 
On the contrary, the relativistic functionals with density-dependent vertex functions 
generally give softer symmetry energies ($50 \leq L \leq 60$ MeV) but PKDD which 
gives $L \approx 90$ MeV.
 The PREX result
yields a very large  central value for $L$, i.e., $L \approx 150$ MeV, but, owing to the 
very large error bars, it constraints very mildly $L$, and  all
theoretical models are compatible with PREX. Other neutron skin measurements
suggest a  smaller range of uncertainties for $L$, but they still have 
non-negligible uncertainties. For instance, the Mainz $(\gamma , \pi^0)$ experiment 
suggests $5 \leq L \leq 55$ MeV. It would be very important
to reduce the experimental uncertainties to obtain more stringent constraints 
on $L$. 
The updated run PREX-II \cite{prex2} aims at a new determination of $R_{skin}$ with
an accuracy of $\pm 0.06$ fm and thus it will constrain the range
of uncertainties of the slope $L$ to $\approx \pm 40$ MeV.

\begin{figure}
	\centering \vskip 1mm
		\includegraphics[scale=0.34, bb= -12 22 720 527, clip]{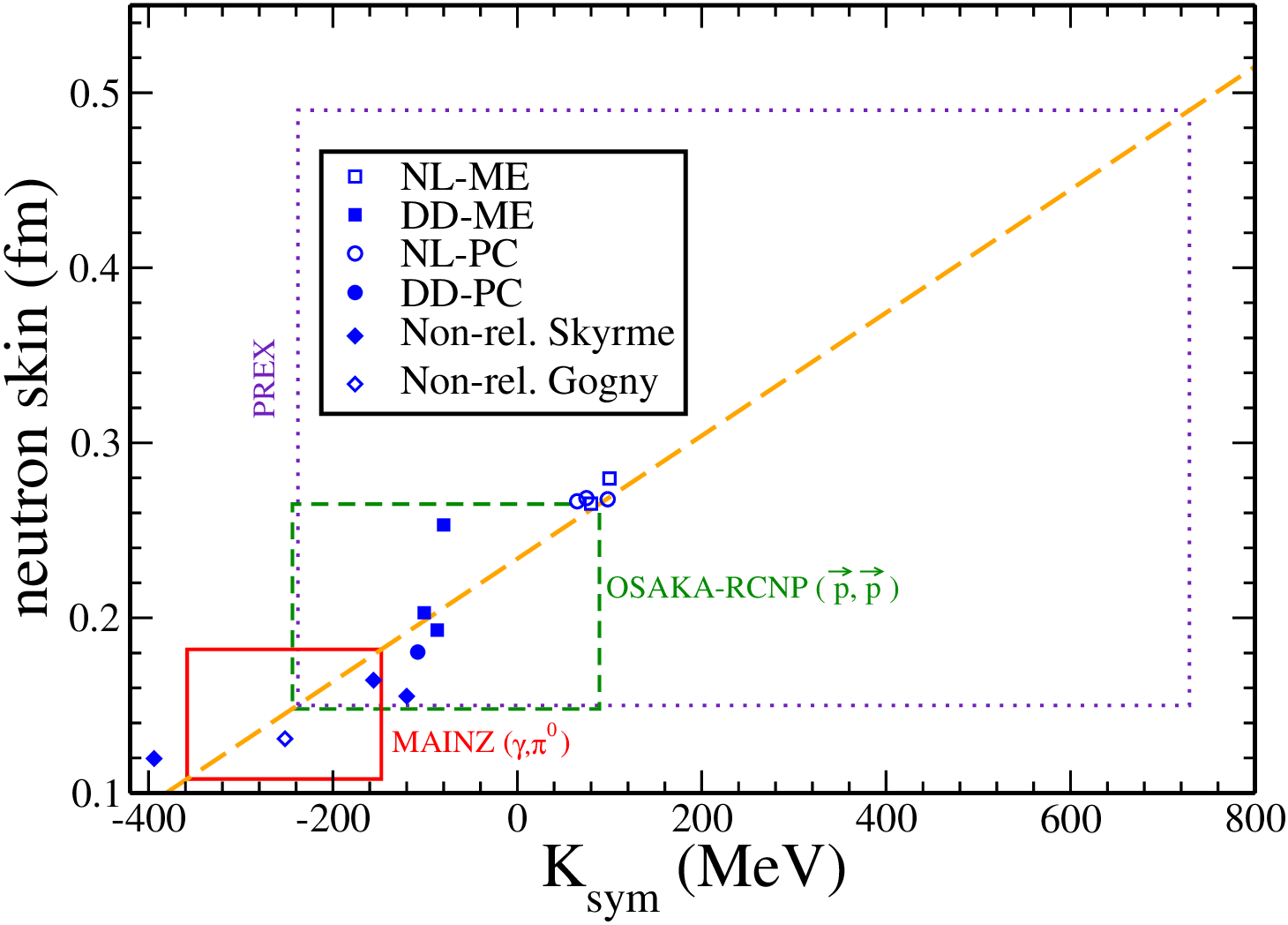}
\caption{ (Color online) 
Neutron skin for $^{208}$Pb versus
 the curvature of the symmetry energy $K_{sym}$ (the orange dashed line represents the best fit correlation). The purple dotted lines are the 
 $1 \sigma$ contours extacted from PREX 
 \cite{Abrahamyan:2012gp}. The red solid lines show the constraints
 from Mainz $(\gamma , \pi^{0})$ measurements \cite{mainz-neutronskin} and
 the green dotted lines  show the constraints on $R_{skin}$
 from Osaka polarized proton elastic scattering measurements \cite{PhysRevC.82.044611}.
}	
\label{fig:k_skin}
\end{figure}

In Fig. \ref{fig:k_skin} we present the correlation of the neutron skin thickness of
 $^{208}$Pb versus $K_{sym}$. In this case the correlation is less strong but it is still
 significant \cite{PhysRevC.72.064309}.  The Gogny and
Skyrme interactions have large negative curvatures $(K_{sym} \le -100$ MeV), while 
most relativistic nuclear interactions lead to positive curvatures ($K_{sym}
\approx 100$ MeV),
but the relativistic functionals with density-dependent vertices give also negative $K_{sym}$.
Very mild constraints on $K_{sym}$ are provided by the PREX result and only the future 
PREX-II \cite{prex2} experiment will constrain the range
of uncertainties of $K_{sym}$ to $\approx \pm 150$ MeV, i.e.,  a constraint  
similar  to that of the Osaka experiment on proton elastic scattering.



\section{Summary and conclusions}

We have presented and discussed numerical predictions for the neutron
density distribution of $^{208}$Pb.
The determination of the neutron distribution in nuclei has proven to be a 
serious challange to our understanding of nuclear structure and it is 
one of the major topics of interest in nuclear physics. Great experimental and
theoretical efforts have been devoted over the last years to achieve this goal. 
In the next years several experiments are planned, in different laboratories 
worldwide, to measure the neutron skin thickness, i.e., the difference 
between the neutron and proton distributions, as accurately as possible. 

Parity-violating electron scattering is an accurate and almost 
model-independent tool for probing neutron properties as it is directly 
related to the Fourier transform of the neutron density.  Starting from various
different theoretical models for nuclear structure, we have extracted the  2pF 
parameters for the neutron distribution and we have compared them with the very recent
 $(\gamma , \pi^0)$ data from  Mainz.
We have then analyzed the linear correlation between the neutron radius and 
the parity-violating asimmetry. The PREX data at average momentum transfer  
$q = 0.475$ fm$^{-1}$ have unfortunately a too much large experimental 
uncertainty to discriminate among the models and only the future run PREX-II
will help to rule out some of them. 
Taking advantage of the linear relations between the neutron skin
 and the slope and the curvature 
 of the symmetry energy around saturation density we can estimate the range
 of variation of $L$ and $K_{sym}$ using the available 
experimental values.

%
\end{document}